\documentclass[aps,prb,amsmath,amssymb,twocolumn,superscriptaddress,showpacs,floatfix]{revtex4-1}

\usepackage{ifpdf}
 \newif\ifpdf
\ifx\pdfoutput\undefined
   \pdffalse
\else
   \pdfoutput=1
   \pdftrue
\fi
\ifpdf
   \usepackage{graphicx}
   \usepackage{epstopdf}
\else
   \usepackage{graphicx}
\fi

\usepackage{bm}
\usepackage{epsfig}
\usepackage[usenames]{color}
\usepackage{soul}
\usepackage{graphicx}
\usepackage{hyperref}

\DeclareMathOperator{\sgn}{sign}

\DeclareMathOperator{\C}{\mathcal C}
\DeclareMathOperator{\Cc}{\mathrm{\scriptscriptstyle C}}
\newcommand{\av}[1]{\left\langle #1\right\rangle}

\newcommand{\Peff}{\vec{P}_{\rm eff}}
\newcommand{\Ceff}{\C_{\rm eff}}

\renewcommand{\paragraph}[1]{\textit{#1.---} } 

\begin{document}

\date{\today }

\title{Interplay of Coulomb Blockade and Ferroelectricity in Nano-Granular Materials}

\author{O.~G.~Udalov}
\affiliation{Department of Physics and Astronomy, California State University Northridge, Northridge, CA 91330, USA}
\affiliation{Institute for Physics of Microstructures, Russian Academy of Science, Nizhny Novgorod, 603950, Russia}

\author{N.~M.~Chtchelkatchev}
\affiliation{Department of Physics and Astronomy, California State University Northridge, Northridge, CA 91330, USA}
\affiliation{L.D. Landau Institute for Theoretical Physics, Russian Academy of Sciences,117940 Moscow, Russia}
\affiliation{Department of Theoretical Physics, Moscow Institute of Physics and Technology, 141700 Moscow, Russia}

\author{A.~Glatz}
\affiliation{Materials Science Division, Argonne National Laboratory, Argonne, Illinois 60439, USA}
\affiliation{Department of Physics, Northern Illinois University, DeKalb, Illinois 60115, USA}

\author{I.~S.~Beloborodov}
\affiliation{Department of Physics and Astronomy, California State University Northridge, Northridge, CA 91330, USA}

\date{\today}

\begin{abstract}
We study electron transport properties of composite ferroelectrics --- materials consisting of metallic grains embedded in a ferroelectric matrix. In particular, we calculate the conductivity in a wide range of temperatures and electric fields, showing pronounced hysteretic behavior. In weak fields, electron cotunneling is the main transport mechanism. In this case, we show that the ferroelectric matrix strongly influences the transport properties through two effects: i) the dependence of the Coulomb gap on the dielectric permittivity of the ferroelectric matrix, which in turn is controlled by temperature and external field; and ii) the dependence of the tunneling matrix elements on the electric polarization of the ferroelectric matrix, which can be tuned by temperature and applied electric field as well. In the case of strong electric fields, the Coulomb gap is suppressed and only the second mechanism is important.
Our results are important for i) thermometers for precise temperature measurements and ii) ferrroelectric memristors.
\end{abstract}


\maketitle

\section{Introduction}\label{sec.intro}

In the past years, composite materials, consisting of conductive grains embedded into some insulating matrix, have attracted continuously increasing attention due to the possibility to combine different, and sometimes competing physical phenomena in a single material and observe new fundamental effects~\cite{Dawber2005RevModPhys,Bel2007review}.
The possible range of observable behaviors is very broad and the following examples by no means exhaustive: granular metals can show the insulator-superconductor transition~\cite{Deut1983,Vill2004,Efetov2002} due to an interplay of superconductivity and Coulomb blockade; or giant magnetoresistance effects appear in granular ferromagnets~\cite{Bel2007,Cir2012} because of the spin dependent tunneling of current carriers between grains; or the
combination of ferroelectric and ferromagnetic materials allows to produce a strain mediated magnetoelectric coupling~\cite{Noh2006,Liu2005,Zhou2007}.

\begin{figure}[t]
\includegraphics[width=2.3in, keepaspectratio=true]{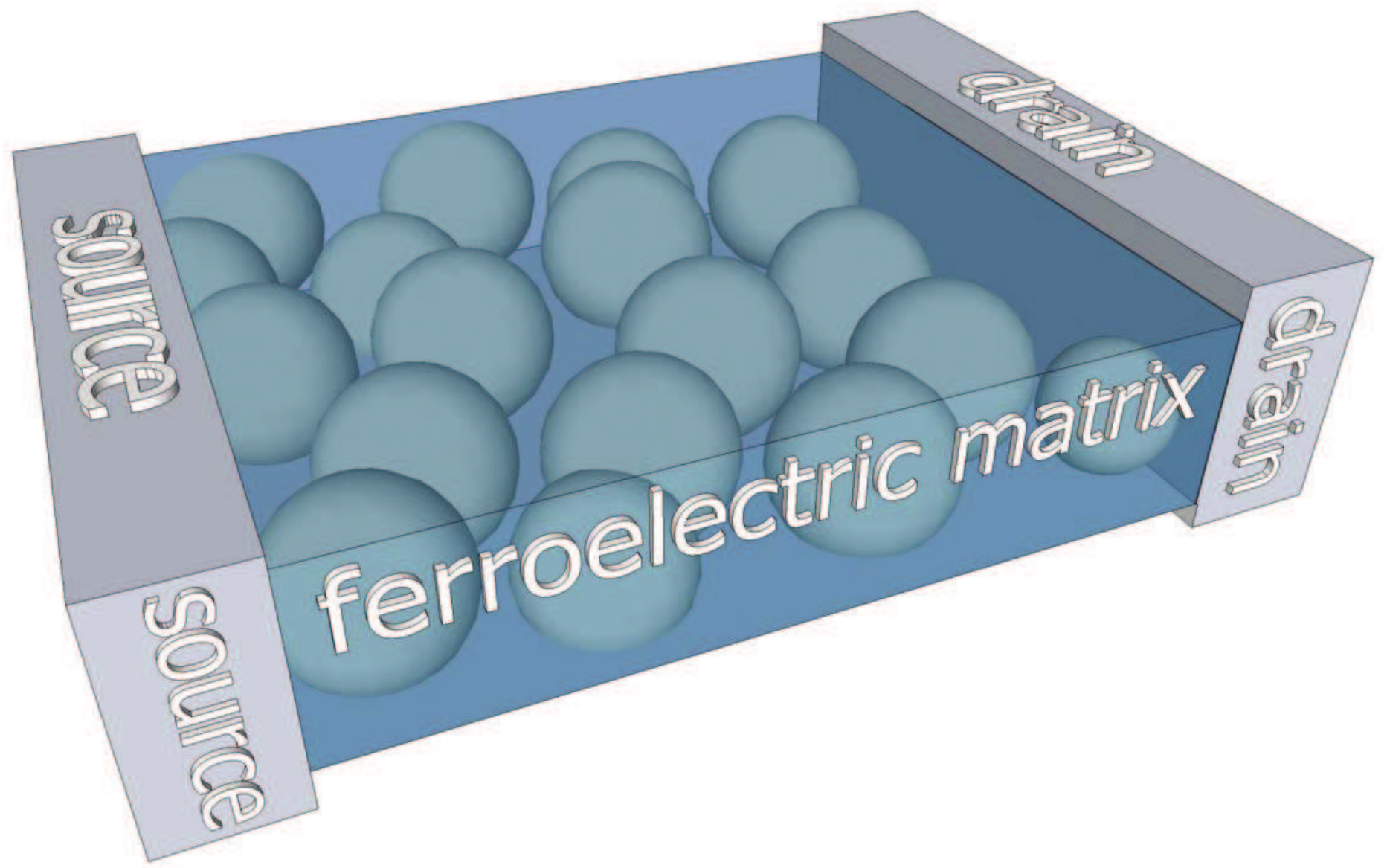}\\
\includegraphics[width=2.3in, keepaspectratio=true]{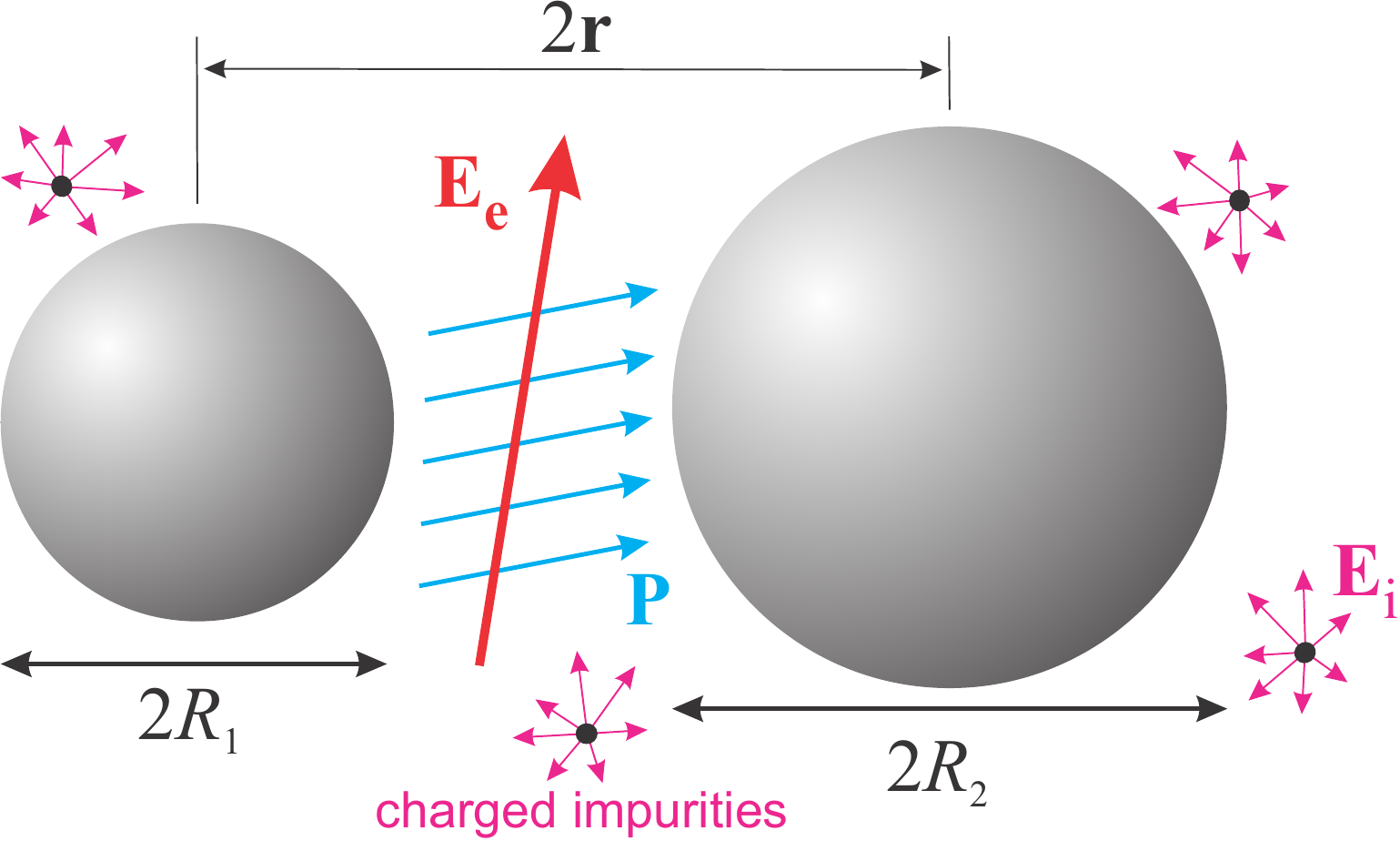}
\caption{\label{Fig_1} (color online) \textit{Top:} Sketch of a granular ferroelectric (GFE)
material with two metal contacts (source \& drain).
\textit{Bottom:} Sketch of a pair of grains embedded in a FE matrix with radii $R_1$ and $R_2$ and distance $2r$ between
them. The vector $\vec{P}$ is the local electric polarization of the FE matrix
and the vector $\vec{E}_i$ is the internal electric field appearing in the system due to the presence
of charged impurities. The vector $\vec{E}_e$ is the applied external electric field.
}
\end{figure}

Besides those fundamental properties, composite materials are promising candidates for concrete microelectronics applications. Composite ferromagnets for example, can be used in magnetic field sensors, due to a high sensitivity of their resistance to a magnetic field change. Granular ferroelectrics  -- subject of this work --  are useful in memory~\cite{Liu2011,Dai2013} and capacitor~\cite{Tuan2007,Moya2001} applications because of their hysteresic behavior and their high dielectric permittivity.

The most interesting and complex aspects of these hybrid systems are their electron transport properties. In particular in (nano) granular materials, several fundamental physical phenomena have to be taken into account in order to develop a theory for the electron conductivity. In this respect the most important are Coulomb blockade~\cite{Efetov2003, Vinokur2005,Vin2003}, grain boundaries~\cite{Bel2007review}, and quantum interference effects~\cite{Varlamov2005,Vin2004}. The transport properties of composite systems are determined by i) the material and the morphology of individual grains and ii) the nature of the coupling between grains.
The conducting grains themselves can be made out of metallic~\cite{Vinokur2005}, superconducting~\cite{Deut1983,Vill2004,Efetov2002}, or ferromagnetic~\cite{Bel2007,Cir2012} materials in various sizes and shapes. The effective coupling strength between grains can be controlled by the materials in which the grains are embedded (the matrix), the grain shape and the nearest-neighbor distance distribution. A typical matrix could be an insulator, a semiconductor, or ligants keeping the grains apart.

Most investigations that address composite (nano) materials dealing with transport physics consider the grains themselves and the emerging transport properties of large grain arrays. In contrast, here we investigate the situation, where the most interesting features of the electron transport appear and are controlled by the insulating matrix. In particular, we investigate composite materials consisting of normal metallic grains embedded in a ferroelectric (FE) matrix,~\cite{Noh2006,Liu2005,Zhou2007} see Fig.~\ref{Fig_1}. In the following we refer to these systems as composite ferroelectrics or more precisely as granular ferroelectrics (GFEs).

Recently composite materials and low-dimensional structures based on ferroelectric matrices
attracted a lot of attention, see, e.g. Refs.~\cite{Scott2007,Dawber2012,Dawber2012_1,dawber2003self,wang2007modeling,zhang2007improved,ortega2012relaxor,Dawber2005RevModPhys,Chanthbouala2012}. However, theoretical investigations of the electron transport in GFEs was limited to the case of rather large intergrain distances, where
the GFEs are practically insulators. It was shown that in the limit of weak external electric fields the conductivity of GFEs strongly depends on the correlation function of the local polarization and the microscopic internal electric field.~\cite{EPL_paper}

In this paper we study electron transport in GFEs in a wide range of external
parameters (electrical field and temperature) that cover not only the insulating,
but also the semiconducting and the metallic regimes. In addition to the dependence of the
tunneling matrix elements on the electric polarization of the ferroelectric matrix,
which can be tuned by temperature and applied electric field~\cite{EPL_paper}, we
also consider two -- so-far unexplored -- effects: i) the dependence of the Coulomb gap on the dielectric permittivity of the
ferroelectric matrix, which in turn is controlled by temperature and external field;
and ii) the hysteresis behavior of the ferroelectric matrix.

Recently, transport properties of composite ferroelectric materials  were studied experimentally~\cite{Dai2013}.
It was shown that GFEs exhibit two important features: i) switching between different resistive states and ii) a current voltage hysteresis.
At the end of this work, we will discuss these experimental findings based
on our theoretical results.

In the out-of-equilibrium regime GFEs are particularly interesting for applications and we will discuss the
possibility to use these materials as memristors (see e.g.~Ref.~\onlinecite{Chanthbouala2012} and Refs. therein).

The paper is organized as follows: In section~\ref{sec:model} we introduce the model of GFEs and calculate the thermodynamic properties of GFEs close to the transition point. In Section~\ref{sec:transport} we study transport properties of GFEs. We discuss our results in Section~\ref{discussions}. Our summary is given
in the conclusion section~\ref{sec:Conclusion}. Finally, we present some estimates for typical materials and discuss the applicability of our results in Appendix~\ref{ApA}.

\section{Microscopic model of composite ferroelectrics\label{sec:model}}

\subsection{Model}
\label{model}

Let us start with our model for composite ferroelectrics. An important feature of GFEs is
the electrostatic disorder in the system. This disorder has two origins:
i) a spatially dependent local anisotropy induced by the grain boundaries; and ii) a strongly inhomogeneous
microscopic internal electric field, $\vec{E}_{i}$. This internal field, generated by charged impurities, see Fig.~\ref{Fig_1}, is effectively screened and its magnitude between two particular grains is defined by the closest impurity located in the FE matrix~\cite{Bel2007review},
$|\vec{E}_{i}|=E_{i}\sim e/(r^2) \sim 10^7-10^9$ V/m with $r$ being the distance from the closest carrier trap, which is of order of a few nm.
This field interacts with the ferroelectric matrix influencing the microscopic distribution of the polarization $\vec{P}$ of the
ferroelectric order parameter and leading to spatial fluctuations of the dielectric permittivity of the ferroelectric matrix.
In addition to the internal, $\vec{E}_{i}$, and external, $\vec{E}_{e}$, fields, the temperature, $T$, also
influences the microscopic structure of the polarization and the dielectric constant.

Granularity introduces additional energy parameters into the problem~\cite{Bel2007review}: each nanoscale cluster
is characterized by (i) the charging energy $E_C = e^2/(\epsilon a)$, where $e$ is the electron charge, $\epsilon$ the dielectric constant,
and $a$ the granule size, and (ii) the mean energy level spacing $\delta$. The charging energy associated
with nanoscale grains can be as large as several hundred Kelvins and we require that $E_c/\delta \gg 1$.
This condition defines the lower limit for the grain size: $a_l = (\epsilon /e^2\nu)^{1/(D-1)}$,
where $\nu$ is the total density of states at the Fermi surface (DOS) and $D$ the grain dimensionality.

The internal conductance of a metallic grain is much larger than the inter-grain tunneling
conductance, which is a standard condition for granularity. The tunneling conductance is one of
the main parameters that controls the macroscopic transport properties of the sample~\cite{Bel2007review}.

The most active regions in the FE matrix are those with the smallest distance between neighboring grains
where electrons can tunnel, see Fig.~\ref{Fig_1}. We describe these regions
as quasi-two dimensional flat interfaces. In composite materials each grain has several neighbors and we enumerate
different pairs of grains (not the grains themselves) by index $i$. Each pair of grains
is characterized by its interface normal
${\vec{n}_{i}} = \vec{r}_i/|\vec{r}_i|$, where $\vec{r}_i$ is the
vector connecting two grains. For grains of equal sizes there is no preferable direction (sign)
of the local normal $\vec{n}_i$. Therefore, we assume without loss of generality that, $(\vec{n}_i\cdot \vec{x}_0)>0$
where $\vec{E}_{e} = E_{e}\vec{x}_0$ with $\vec{x}_0$ being the direction of $x$-axis.
This condition defines the direction of vector $\vec{n}_i$.

For two-dimensional interfaces the electric polarization is perpendicular to the interface, i.e.,
directed along the surface normal. Since the correlation length of the ferroelectric order parameter
can be of the order of $1$ nm for temperatures not very close to the critical
temperature we can assume that the local polarization $\vec{P}$ follows
the local normal vectors $\vec{n}_i$, see Appendix A for details.

We also assume that external, $\vec{E}_{e}$ and internal, $\vec{E}_{i}$ electric fields
do not change the orientation of the polarization
(only the sign of the polarization can be changed by the electric field).
We describe the internal and external electric fields by two
angles $\theta_{i,i}$ and $\theta_{e,i}$ with respect to the normal $\vec{n}_i$.

Next we discuss important thermodynamic characteristics controlling the electron transport in GFE.
First, we discuss the properties of local polarization and susceptibility concentrating on
a single ferroelectric interlayer between pair of grains. Second, we consider average
quantities of GFE. Finally, we discuss the hysteresis phenomena appearing in GFE.

\subsection{Local polarization and susceptibility}

Next, we discuss the properties of local polarization and susceptibility.
To simplify our notations we will omit the grain pair index $i$ in the following.

The local polarization $\vec{P}$ can be written as $\vec{P}=P(E_{\vec{n}})\vec{n}$ with $E_{\vec{n}}=E_{e}\cos(\theta_{e})+E_{i}\cos(\theta_{i})$. To describe the polarization $P(E_{\vec{n}})$ we use the Landau-Ginzburg-Devonshire theory~\cite{devonshire1949xcvi,falk1980model,Levan1983,Tilley2001,chandra2007landau} with the free energy density written in the form
\begin{equation} \label{Eq_1}
F = F_{0} + \alpha P^2+\beta P^4-E_{\vec{n}}P.
\end{equation}
Here $F_0$ is the polarization independent part of the free energy density. The validity of the mean field theory in thin ferroelectrics
is discussed in the Appendix~\ref{ApA}. Close to the transition temperature $T_{\Cc}$ the parameter $\alpha$ has the form $\alpha = \eta(T-T_{\Cc})$, and $\beta$ does not depend on temperature~\cite{Levan1983}. Equation~(\ref{Eq_1}) does not take into
account the non-uniformity of the polarization $P$.
All transport characteristics for an arbitrary FE can be obtained if the function $P(E_{\vec{n}})$ is known.

Above the transition temperature $T_{\Cc}$ a non-zero polarization $P$ appears only for a finite electric field. Below $T_{\Cc}$
a spontaneous polarization appears even for zero electric field.

A hysteresis loop exists, as usual, only  below the transition temperature $T_{\Cc}$. Switching between two branches of hysteresis loop occurs at the switching field $E_s=4\alpha\sqrt{|\alpha|/6\beta}/3$. 

The local dielectric susceptibility along the
direction $\vec{n}$ is given by:~\cite{Levan1983} $\chi_{\vec{n}}=\partial_{E_{\vec{n}}} P=(2\alpha+12\beta P^2(E_{\vec{n}}))^{-1}$. For temperatures below the transition temperature, $T < T_{\Cc}$, the dielectric susceptibility $\chi_{\vec{n}}$ diverges at the points of the polarization switching $E_{\vec{n}} = \pm E_{s}$. For temperatures above $T_{\Cc}$ the permittivity $\chi_{\vec{n}}$ is a smooth function without any singularities. The above discussions are valid for local properties of composite materials only.

\subsection{Macroscopic susceptibility and correlation function}

The electron transport in GFE is controlled by the average electrical susceptibility $\overline{\chi}$ and
the correlation function of local electric field and local polarization of FE matrix $C=\av{(\vec{E}_i+\vec{E}_e)\cdot\vec{P}}$.  Using the microscopic model of FE matrix discussed in Sec.~\ref{model}
we calculate $\overline{\chi}$ and $C$ by averaging over the
mutual orientation of local normal $\vec{n}$, internal $\vec{E}_i$, and external $\vec{E}_e$ electric fields.
The details of these calculations are relocated into Appendix B.

\begin{figure}[t]
\includegraphics[width=3.25in, keepaspectratio=true]{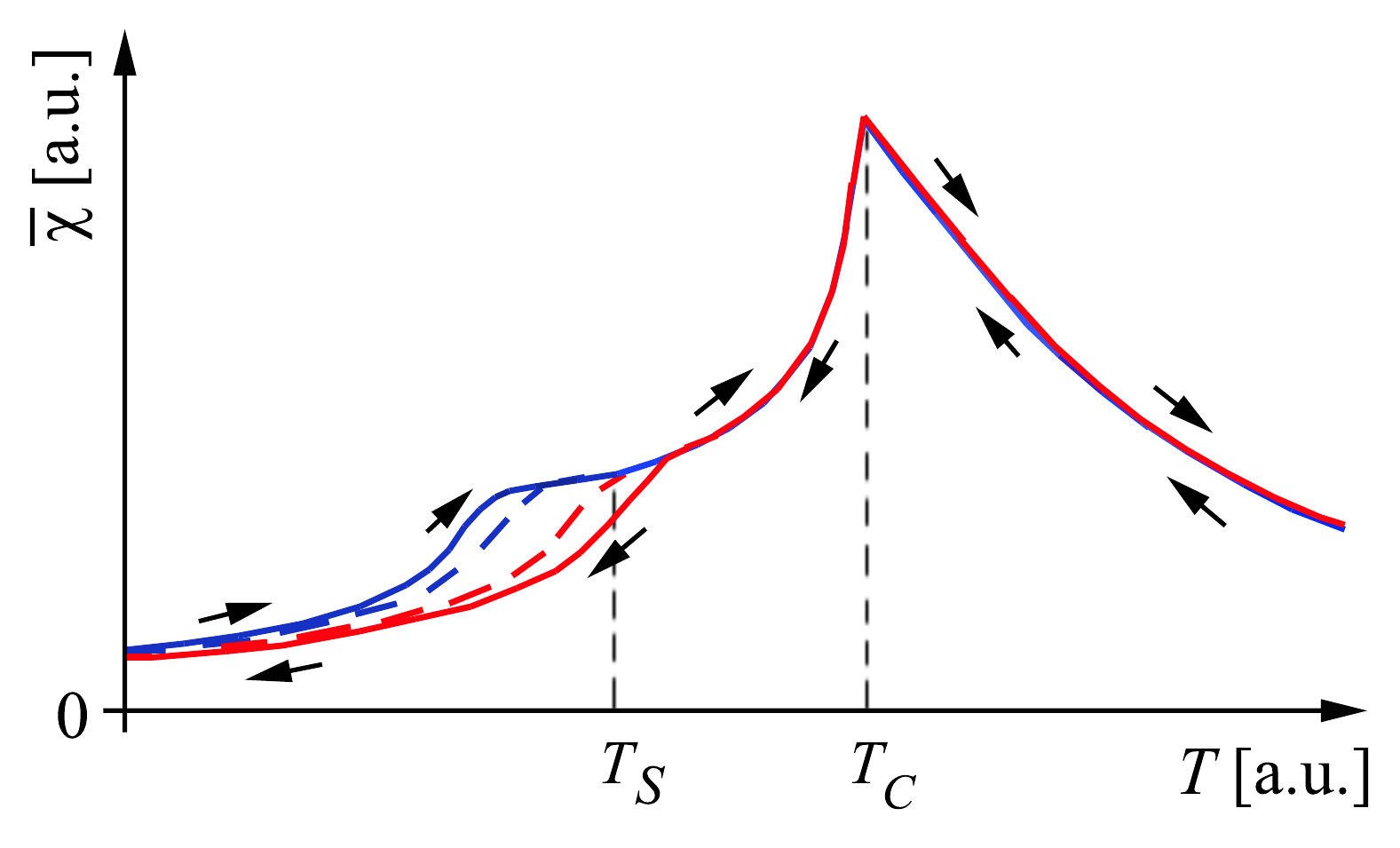}%
\caption{\label{Fig_SusT} (color online) Average dielectric susceptibility $\overline{\chi}$ vs. temperature.
The solid lines correspond to the longitudinal $\overline{\chi}_{\parallel}$, Eq.~(\ref{Eq_11}), and the dash
lines correspond to the perpendicular $\overline{\chi}_{\perp}$, Eq.~(\ref{Eq_12}),
components of susceptibility $\overline{\chi}$.  The behavior is shown for
both hysteresis branches. Arrows indicate the path around the hysteresis loop for
fixed external electric field $E_e = E_i/3$. $T_C$ and $T_S$ are the Curie and the switching temperatures, respectively.
The derivative of susceptibility has a jump at the transition point following from the divergence
of the microscopic susceptibility $\chi_{\vec{n}}$ appearing at the points of
polarization switching ($E_s$ or $T_S$). However, in real ferroelectrics this
behavior is absent due to order parameter fluctuations.
Therefore the kink in the average susceptibility $\overline{\chi}$ is smeared.}
\end{figure}

The final result for temperature dependence of dielectric susceptibility in Eqs.~(\ref{Eq_11}) and Eq.~(\ref{Eq_12})
is shown in Fig.~\ref{Fig_SusT}. The susceptibility has its maximum
value  $\overline{\chi}_{\|}=\overline{\chi}_{\perp}=4^{2/3}/(24\beta^{1/3}E_i^{2/3})$ for temperature $T=T_{\Cc}$
and zero external field, $E_e=0$, and it increases with decreasing internal field. The
derivative of susceptibility has a jump at the transition point following from the divergence
of the microscopic susceptibility $\chi_{\vec{n}}$ appearing at the Curie temperature $T_{\Cc}$. However, in real ferroelectrics this behavior is absent due to order parameter fluctuations.
Therefore the kink in the average susceptibility $\overline{\chi}$ is smeared.
There is also another peculiarity in the temperature dependence of the susceptibility $\overline{\chi}$.
at finite external electric field $E_e$ due to the hysteresis behavior of the polarization of FE matrix.
This peculiarity is located at the switching temperature $T_S$, defined by the equation
$E_s(T_S) = <|E_{\vec{n}}|>$, with $E_s$ being the switching field and brackets standing for
averaging over all pairs of grains.

The behavior of the correlation function $C$ is shown in the inset
in Fig.~\ref{Fig_Met}. It has two branches below the critical temperature $T_C$
due to hysteresis behavior of local polarization. The upper branch corresponds to the
case of local polarization directed along the applied electric field, in this case the correlation
function is positive, $C>0$. The lower branch
corresponds to the situation with local polarization directed oppositely to the
electrical field, in this case $C<0$. Above the Curie temperature $T_C$ the
correlation function $C$ has only one branch and monotonically decreases with increasing the temperature $T$.

\subsection{Hysteresis}

The properties of GFE depends on it's history. To study the hysteresis phenomena, we first apply a large positive external electric field ($E_{e}>0$) and then decrease its magnitude until reaching a large negative field ($E_{e}<0$) [upper branch], which is finally reversed until the initial electric field value is reached [lower branch], thus closing the hysteresis loop. As a result the temperature dependence has two branches.

\section{Electron transport in composite ferroelectrics}
\label{sec:transport}

In this section we discuss the transport properties of composite ferroelectrics.
The ferroelectric matrix influences the transport properties
in two ways:

i) Through the dependence of the local tunneling conductance
between two grains $\tilde{g}_{t}$ on the polarization, $\tilde{g}_{t}=g^{0}_{t}(1 + \zeta ((\vec{E}_{i}+\vec{E}_{e})\cdot\vec{P}) + \mu ((\vec{E}_{i}+\vec{E}_{e})\cdot\vec{r})(\vec{P}\cdot\vec{r}))$ with $g_t^0$ being the
tunneling conductance in the paraelectric state and $\zeta$, $\mu$ being phenomenological constants,
and $\vec{r}$ being the vector connecting the grains (see Fig.~\ref{Fig_1}).

ii) Through the dependence of Coulomb gap
\begin{equation}
\label{Eq_Gap}
E_C = e^2/(\epsilon a).
\end{equation}
The Coulomb blockade leads to the appearance of the Mott gap, $E_C$,
allowing for an additional control of the transport properties by external
electric field and temperature.
This effect is especially pronounced for weak external
electric fields and low temperatures, where transport is due to electron cotunneling~\cite{Bel2007review}.
For strong external fields the Coulomb blockade is suppressed, thus leading to a weak
dependence of the conductivity on the dielectric permittivity.

There are several transport regimes in composite materials depending on the coupling between the grains.
For weak coupling, low temperatures, and small electric fields the electron transport is
due to electron cotunneling. This mechanism involves electron energy levels inside the Mott gap.
At higher temperatures electrons can be excited directly above the Coulomb gap. Thus,
the activation transport mechanism becomes important. At even higher temperatures, $T \geq E_C$,
the electron transport becomes metallic.

We mention that for temperatures approaching the transition temperature $T_{\Cc}$ the dielectric permittivity $\epsilon$
increases leading to a decrease of the charging energy $E_C = e^2/\epsilon a$, with
$\epsilon$ being the permittivity of the whole sample including the ferroelectric matrix and metallic grains.
Assuming that the metal dielectric
constant is very large (infinite) at zero frequency we can write for sample permittivity
\begin{equation}\label{Eq_eps}
\epsilon = \epsilon_{fe} (\Omega/\Omega_{fe}), \hspace{0.5cm} \epsilon_{fe}=1+4\pi\overline{\chi},
\end{equation}
were $\Omega$ and $\Omega_{fe}$ are
the sample and ferroelectric matrix volume, respectively.

The value of the dielectric permittivity, $\epsilon_A$, where activation transport becomes
important is $\epsilon_A=e^2\xi/(a^2 k_B T)$, where $\xi$ is the electron localization
length defined below~\cite{Bel2007review}. If the maximum
of $\epsilon=1+4\pi\overline{\chi}$ is large at temperatures approaching $T_{\Cc}$, $\epsilon > \epsilon_A$
then one observes activation transport in a temperature region $T^<_A < T < T^>_A$, where temperatures
$T^<_A$ and $T^>_A$ are defined by the condition $\epsilon(T^{<,>}_A)=\epsilon_A(T^{<,>}_A)$.
For grain sizes $a=4$nm and temperature $T\approx 400$ K one finds $\epsilon_A\approx9$.

Metallic regime appears for temperatures $T_M\ge e^2/(\epsilon(T_M)a)$.

Usually ferroelectrics have a very large dielectric constants in the vicinity
of the transition temperature leading to  the merallic transport in this temperature region.

Another important parameter controlling the GFE conductivity is the external electric field, $E_e$.
The phonon mediated electron cotunneling occurs for weak external electric fields,
$E_e < E^* = T/(e \xi) \approx 10^6-10^8$ V/m. For electric fields $E_e > E^*$
the electron cotunneling is mediated by external electric field and does not depend on temperature.
\begin{figure}[t]
\includegraphics[width=3.25in, keepaspectratio=true]{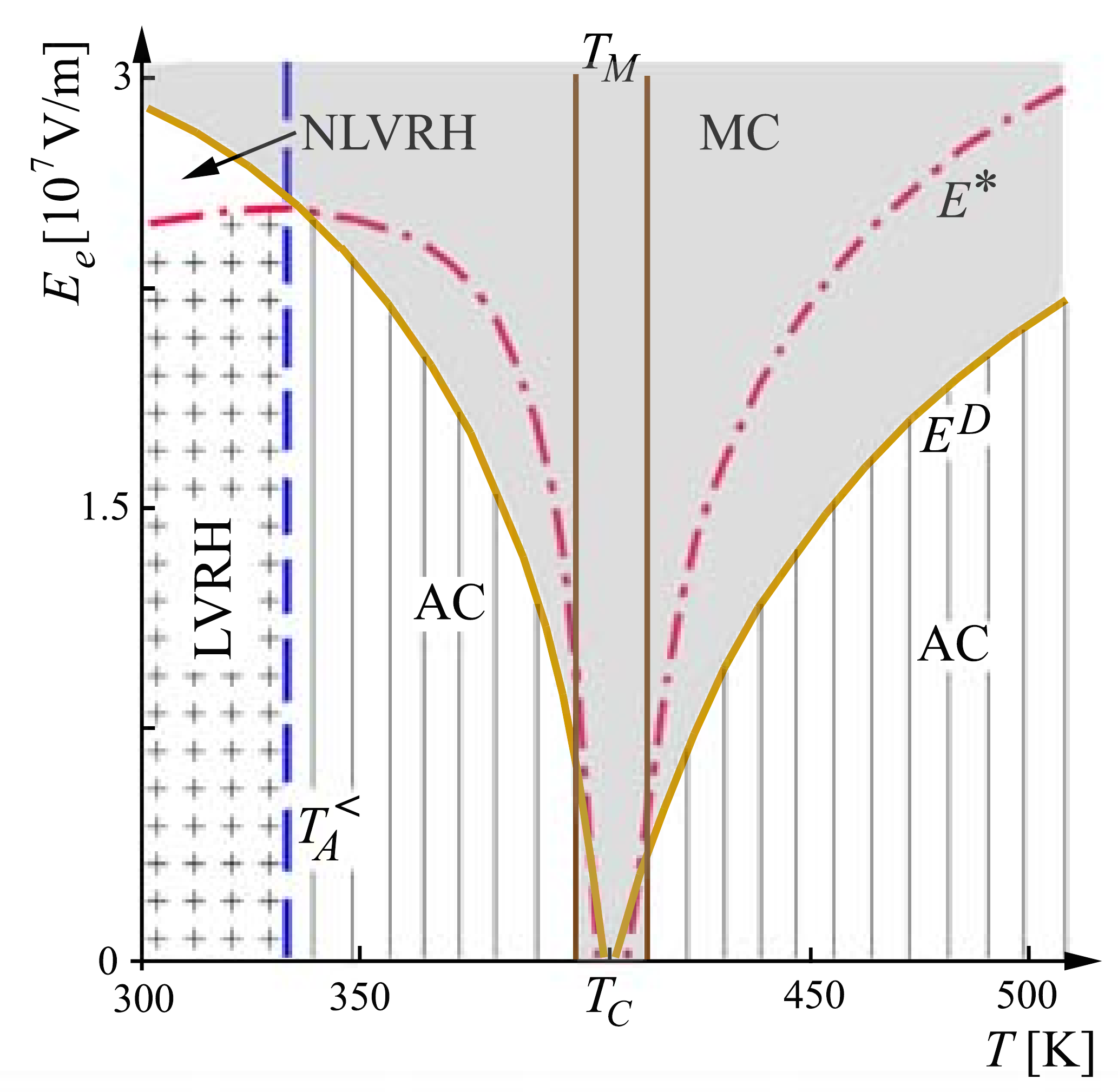}%
\caption{\label{Fig_Diag} (color online) Transport phase diagram of granular ferroelectrics in coordinates of the external
electric field $E_e$ vs. temperature $T$. $T_C$ is the Curie temperature.
AC (shaded region) denotes the activation conductivity, LVRH (cross filled region) and NLVRH (unfilled region)
stand for linear and non-linear electron cotunneling, respectively.
MC (colored region) stands for metallic conductivity. The blue line $T_A^<$ describes the transition to activation transport;
the brown lines $T_M$ describe the transition from activation to metallic regimes, these lines have a physical
meaning outside the metallic regime only (colored region - MC)
The electric field $E^D$ describes the transition to the metallic regime.
The field $E^*$ shows the boundary between the linear and non-linear hopping regimes.
This field has physical meaning outside the metallic regimes only, since inside the metallic
region the VRH contribution to the conductivity is negligible.}
\end{figure}
In addition, there is another
characteristic field $E^D = e/(a^2\epsilon) \approx 10^6-10^8/$ V/m.
For external fields larger
than $E^D$ the electron transport is metallic.
The ratio of the two fields is $E^D/E^*=(E_C/T)\cdot(\xi/a)$.

We summarize the different transport regimes of granular ferroelectrics as a
function of external electric field and temperature in the reversible case
in Fig.~\ref{Fig_Diag} for the following set of parameters:
$T_{\Cc}=400$ K, $\eta=0.01$ 1/K, $\beta=3\cdot10^{-2}$, this corresponds to BaTiO$_3$;
parameter $\zeta=10^{-10}$ is chosen to be small such that hysteresis effects can be neglected;
$g^t_0=0.2$, $a=5$ nm, $\Omega/\Omega_{fe}=1.5$, the last parameter corresponds to a $1$ nm
distance between the grains, and $E_i=7\cdot10^{8}$ V/m.
Below we discuss three transport regimes in more details.

\subsection{Electron cotunneling}

At low external electric fields and weak coupling between the grains the electron transport is due to the cotunneling mechanism.
The most important parameters are the average tunnel conductance $g_t(P)$ and the electron localization length $\xi$~\cite{Bel2007review}. The former is given by
\begin{eqnarray}
g_{t}(P) &=& g^0_t\left(1 + \Ceff \right)\,\,\,\text{with}\label{Eq_19} \\
\Ceff &\equiv& \av{\vec{E}\cdot\Peff}\label{eq.CF}
\end{eqnarray}
being the correlation function of the effective polarization with
$\Peff = \zeta \vec{P}+\mu \vec{r}_{12}(\vec{P}\cdot\vec{r}_{12})$ and the electric field $\vec{E}=\vec{E}_{i}+\vec{E}_e$;
vector $\vec{r}_{12}$ connects two grains;
$g^0_t$ is the tunneling conductance in the paraelectric phase. The inelastic localization length $\xi$ is
given by the expression~\cite{Bel2007review}
\begin{equation} \label{Eq_20}
\xi=a/\ln(E^{2}_{c}/T^2g^{0}_{t}).
\end{equation}
The conductivity in this regime is
\begin{equation} \label{Eq_21}
\sigma_{L} = g^0_t(1 + \Ceff)\exp(-\sqrt{T^P_0/T}),
\end{equation}
where $T^P_0$ is the characteristic temperature scale
\begin{eqnarray} \label{Eq_22}
T^P_0 = T_0 \left[1 - \frac{\xi}{2a} \ln\left(1 +  \Ceff \right) \right],
\end{eqnarray}
with $T_0=e^2/(\epsilon\xi)$, \cite{Sh1975}. To calculate $\Ceff$ one has to evaluate first the average $\av{(\vec{E}\cdot\vec{r}_{12})(\vec{r}_{12}\cdot \vec{P})}$. Using Wick's theorem one can show that $\Ceff=\tilde{\zeta}C$, where $\tilde{\zeta}=\zeta+\mu\av{\vec{r}^2_{12}}$.

In the non-linear (field driven) cotunneling regime with external electric fields
$E_e > E^*$, the conductivity has the following form:
\begin{equation}\label{23}
\sigma_{NL} = g^0_t(1 + \Ceff)\exp(-(E^W_0/E_{e})^{1/2}).
\end{equation}
Here $E^W_0=T^P_0/e\xi$ is the characteristic electric field with temperature $T^P_0$, \cite{Sh1973}.

\subsection{Metallic transport}

For strong external electric field, $E_e > E^D$, the Coulomb blockade is suppressed
leading to metallic transport. The main contribution to conductivity in this
regime is given by the expression~\cite{Bel2007review}
\begin{equation}\label{24}
\sigma_D=2e^2g_t/a=2e^2g^0_t(1 + \Ceff)/a.
\end{equation}

\subsection{Activation transport}

Another regime which is shown in Fig.~\ref{Fig_Diag} is the region with activation
transport where the main contribution to the conductivity is due to
electron driven by the temperature to the conduction band, above the Mott gap
\begin{equation}\label{25}
\sigma_{A} \sim \exp(-[E_C/T]).
\end{equation}

\section{Discussion}
\label{discussions}

\subsection{Metal-insulator transition}

In this section we discuss two transport phenomena specific to granular ferroelectrics
starting our discussions with the metal-insulator transition.

The transport phase diagram in Fig.~\ref{Fig_Diag} shows two transitions
for temperatures close to the critical temperature $T_C$ and weak applied electric fields: i)
from VRH to activation transport and ii) from activation
to a metallic transport. These transitions are possible due to a strong dependence of the
Coulomb gap on temperature.

\begin{figure}[t]
\includegraphics[width=3.25in, keepaspectratio=true]{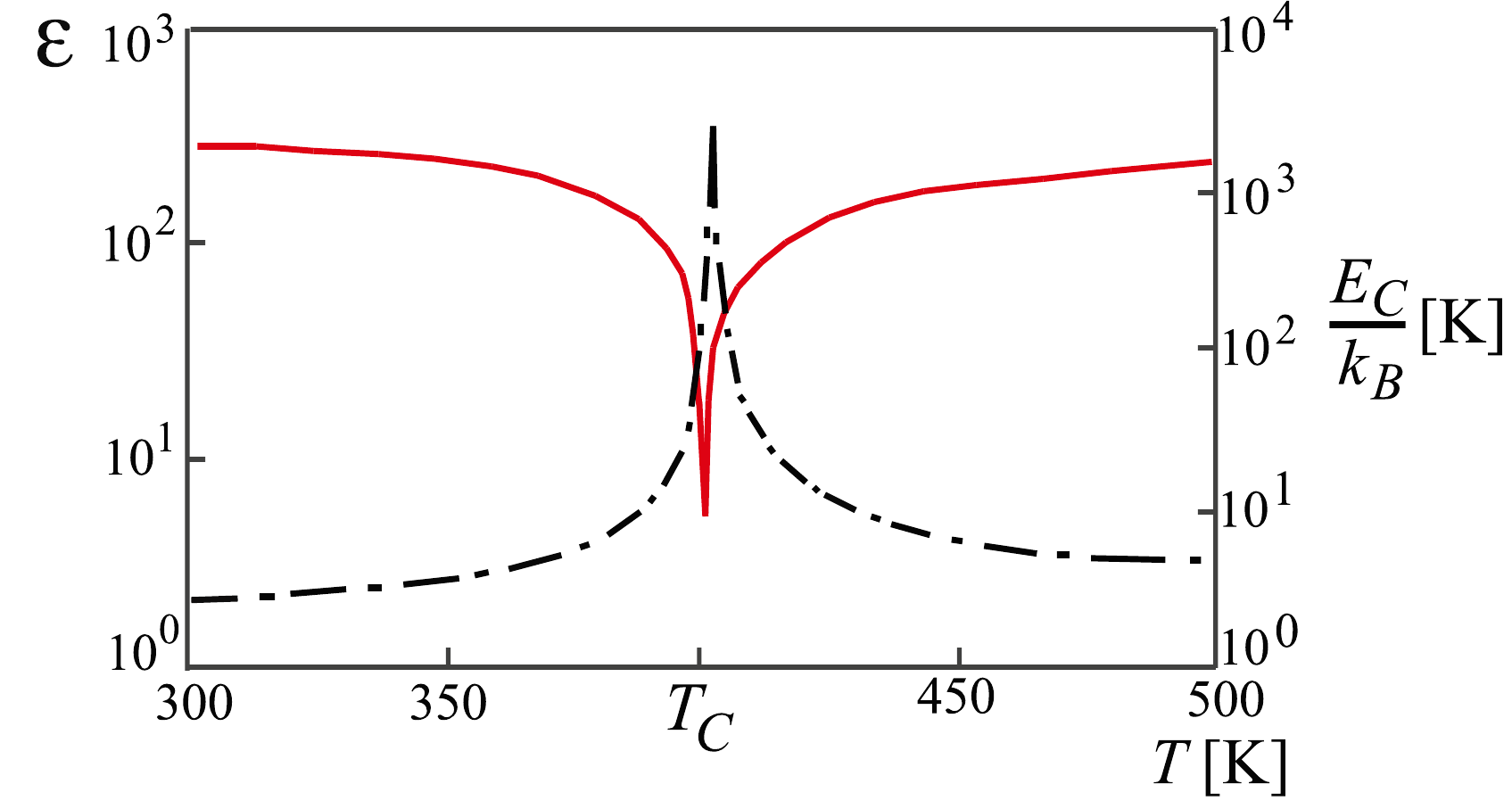}%
\caption{\label{Fig_Perm} (Color online) Dielectric permittivity $\epsilon$ (solid line), Eq.~(\ref{Eq_eps}),
and Coulomb gap $E_C$ (dash line) of granular ferroelectrics, Eq.~(\ref{Eq_Gap}),
vs. temperature at fixed external electric field, $E_{e}= 6\cdot10^6$ V/m. $T_C$ is the Curie temperature.}
\end{figure}

The dependence of the dielectric permittivity and the Coulomb gap on temperature is shown in
Fig.~\ref{Fig_Perm}. The curves are plotted for the same set of parameters as used in the
previous section for the calculation of the transport phase diagram and for external electric field we use
$E_{e}=6\cdot10^6$ V/m. The dielectric permittivity of ferroelectric materials
diverges close to the transition temperature on both sides of the
transition as $\epsilon\sim 1/|(T-T_C)|$.
However, due to the granular morphology and the internal electric field the permittivity peak is smeared.

The transition from VRH to activation conductivity can be understood as follows:
Away from the transition temperature $T_C$
the dielectric permittivity $\epsilon$ of the FE matrix is small and the Coulomb gap is large.
Therefore there are no electrons in the conduction band at zero temperature, thus
the GFE is an insulator. The only transport mechanism here is VRH. The increase of temperature
leads to the reduction of the Coulomb gap and to the increase of the number of electrons in the conduction band
(above the Mott gap). The activation conductivity becomes more important
than VRH at temperatures $T \geq T_M$.

The transition from activation to the metallic regime can
occur in two ways (see Fig.~\ref{Fig_Diag}):
i) the temperature driven, for external fields less than $ E_e < 5\cdot 10^6$V/m.
In this case the Mott gap disappears for temperatures approaching the Curie point leading the GFE
to the metallic state. This transition occurs at temperatures $T = T_M$.

ii) the field driven, for external fields larger than $E_e > 5\cdot 10^6$V/m.
In this case the voltage between the nearest neighbor grains becomes
comparable with the Mott gap pushing electrons to the conduction band. This transition
occurs at external electric field $E_e = E^D$.

\begin{figure}[t]
\includegraphics[width=3.25in, keepaspectratio=true]{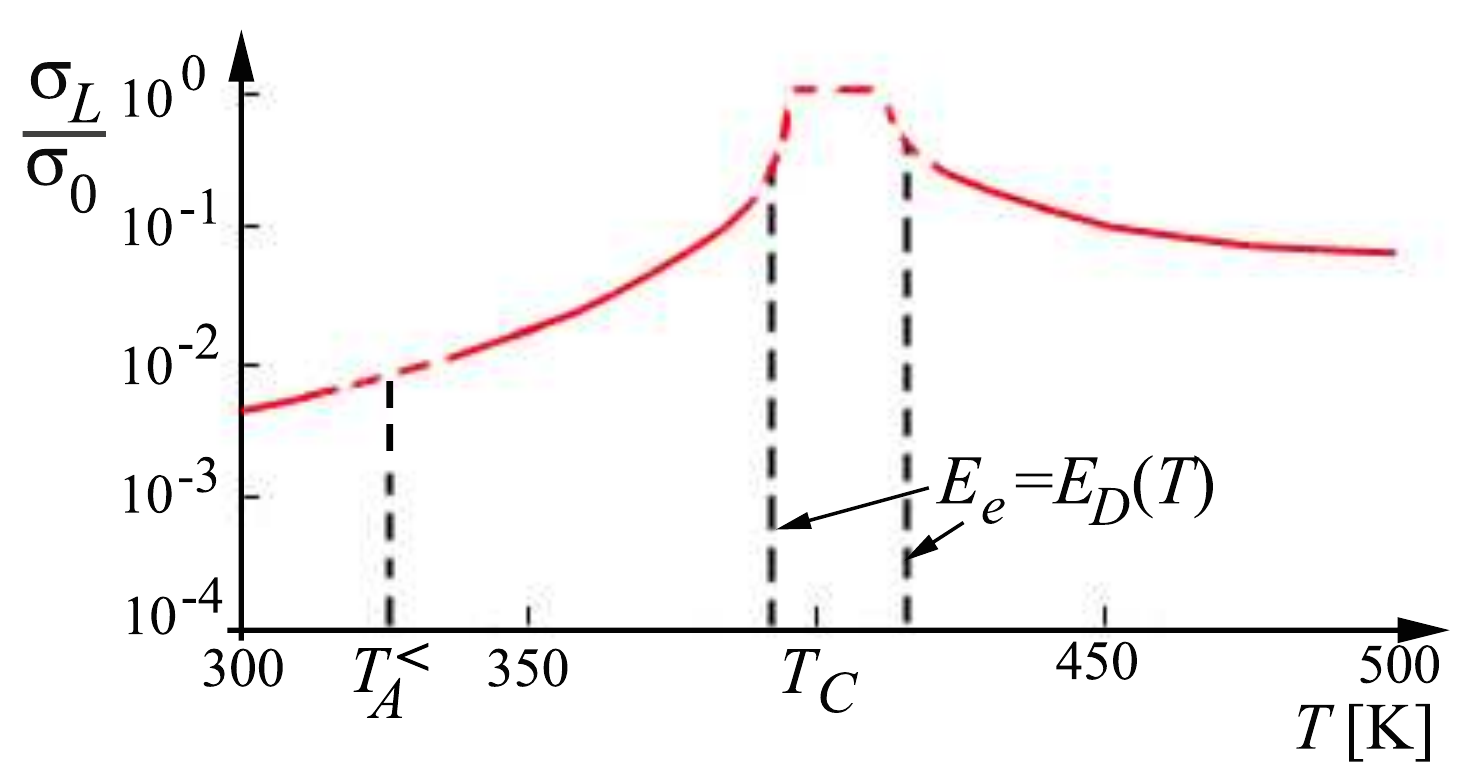}%
\caption{\label{Fig1_MI} (Color online) Conductivity $\sigma$ of GFE, with $\sigma_0=2e^2g^0_t/a$ being the
metallic conductivity in the paraelectric phase, vs. temperature at fixed external electric field $E_e=6\cdot10^6$ V/m.
Variable range hopping and electron cotunneling, Eq.~(\ref{Eq_21}),
is the main transport mechanism for temperatures $T < T_A^<$. Close to the transition temperature
$T_C = 400$ K the transport is metallic, Eq.~(\ref{24}). Between these two regions the
conductivity has activation behavior, Eq.~(\ref{25}).}
\end{figure}

Figure~\ref{Fig1_MI} shows the metal-insulator transition corresponding to the temperature dependence of the
Coulomb gap presented in Fig.~\ref{Fig_Perm}. Figure~\ref{Fig1_MI} is plotted for small
susceptibility $\chi$, where hysteresis effects can be neglected.
For these parameters there are two clear transitions:
i) from VRH to activation transport at temperatures $T = T_A^<$ and ii) from activation to
metallic transport at external electric field $ E_e = E^D$. The field driven transition
occurs when the horizontal line in the transport phase diagram in Fig.~\ref{Fig_Diag}
corresponding to the external field $E_e = 6\cdot10^6$ V/m crosses the $E^D$ curve.

It follows from Fig.~\ref{Fig1_MI} that the conductivity of GFE increases three orders of
magnitude in a rather narrow temperature range. This is an unexpected result because
usually conductivity decreases in the vicinity of a phase transition due to scattering of electrons
on fluctuations of the order parameter. Here we have the opposite situation with
decreasing resistivity. This behavior can be utilized to built a GFE thermometer for precise temperature measurements using an appropriate gauge.
It is worth to mention that this non-trivial behavior is a peculiarity of granular ferroelectrics and cannot be
observed in the tunnel junctions with ferroelectric barrier.

\subsection{Memory effects}

\begin{figure}[t]
\includegraphics[width=3.25in, keepaspectratio=true]{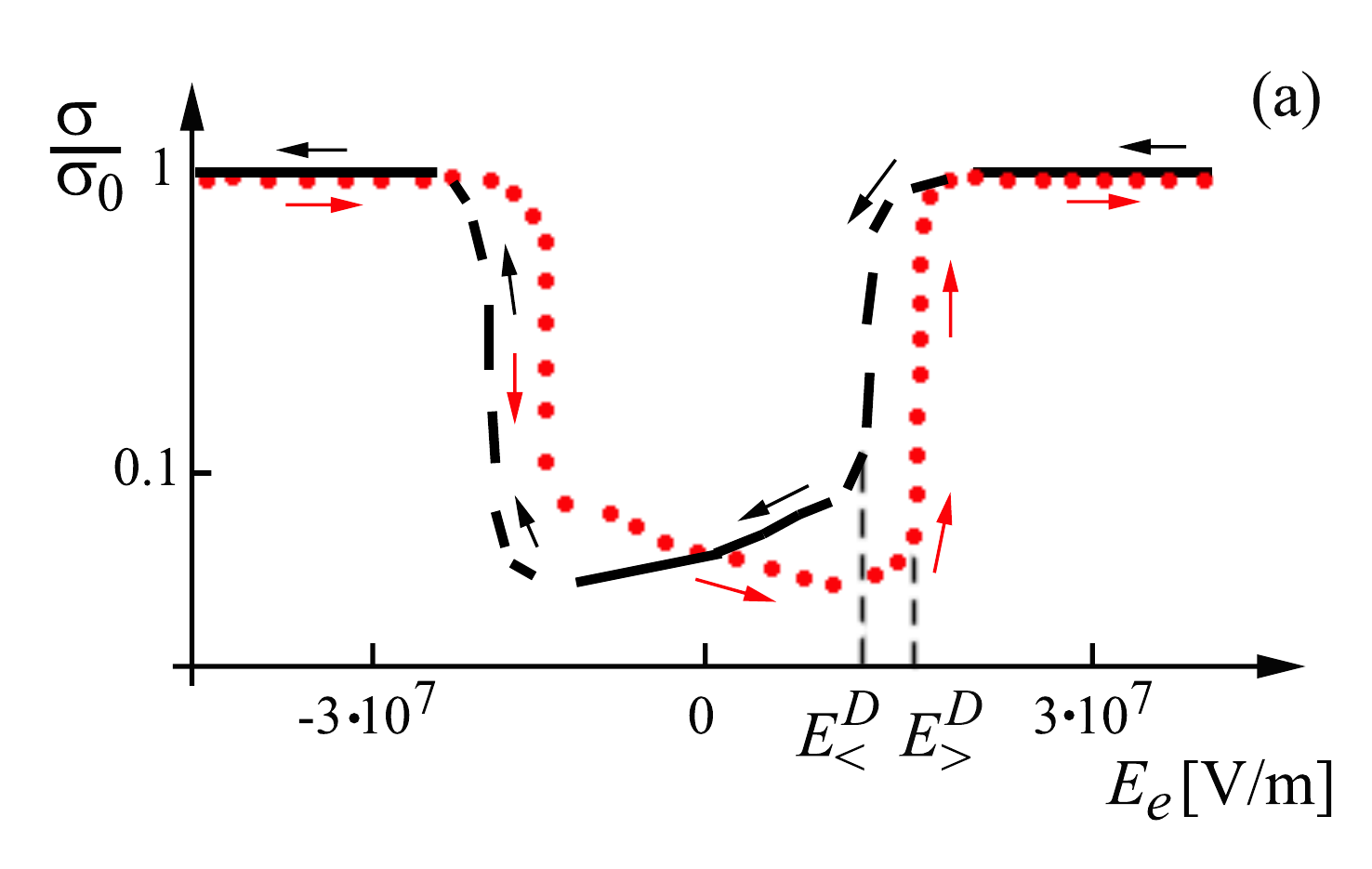}\\
\includegraphics[width=3.25in, keepaspectratio=true]{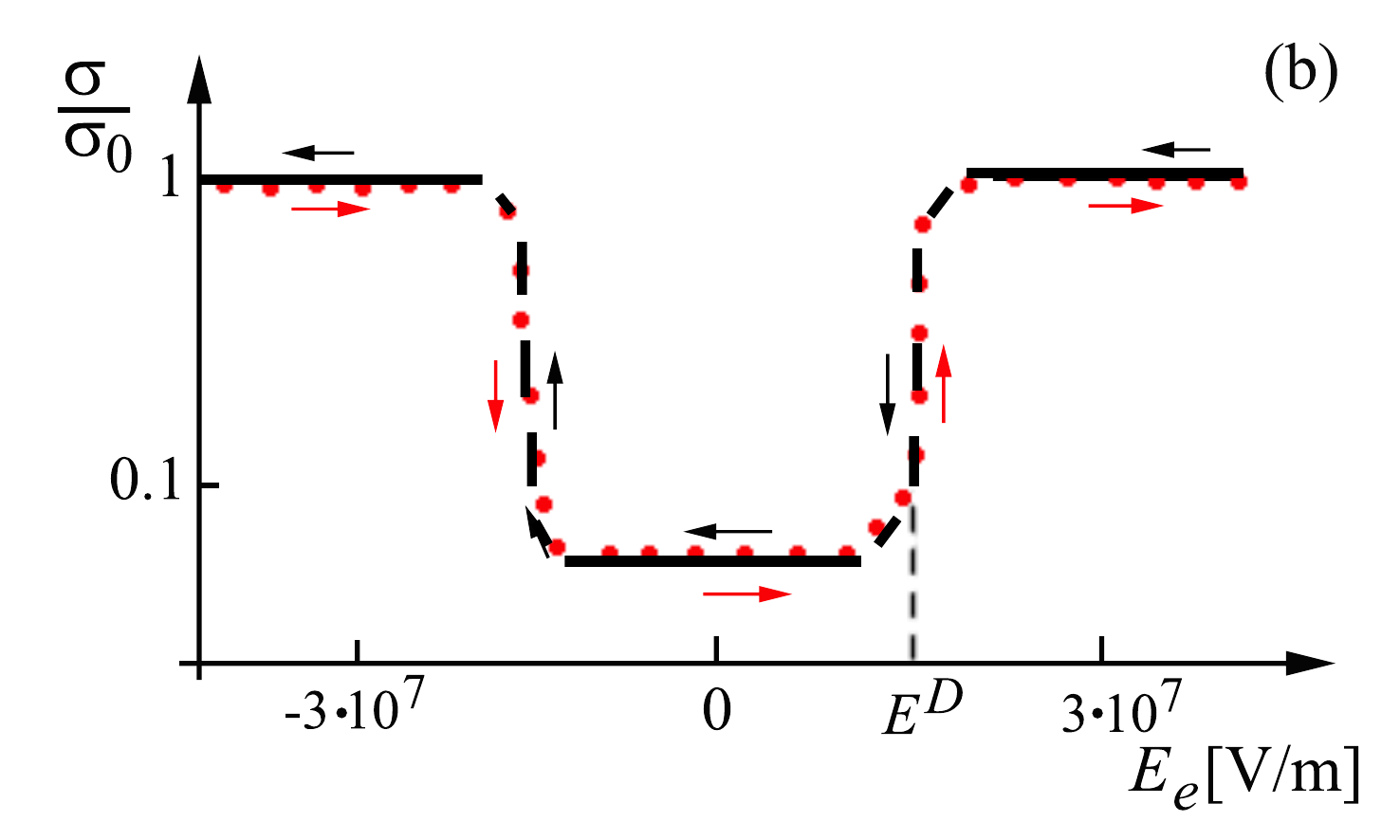}
\caption{\label{Fig_E} (color online) Conductivity $\sigma$, with $\sigma_0=2e^2g^0_t/a$ being the
metallic conductivity in the paraelectric phase, vs. external electric field $E_e$ at fixed temperature $T = 350$ K ($E^D\approx 1.5\cdot10^7$ V/m). The solid and dotted lines correspond to the two hysteresis branches. The arrows show the direction, the hysteresis loop is followed.
There are two different situations: (a) The average internal field $E_i=5.7\cdot10^7$ V/m is less than the FE
switching field $E_s=6.3\cdot10^7$ V/m and (b) The average internal field $E_i=5.7\cdot10^7$ V/m is larger than the
switching field $E_s=3.9\cdot10^7$ V/m. For internal fields $E_i > E_s$,
there is no difference between fields $E^D_<$ and $E^D_>$.}
\end{figure}

In the previous section we discussed the influence of the FE matrix on the Coulomb gap
of the GFE system. Here, on the other hand, we study explicitly the influence of the hysteretic behavior  on the electron transport in GFEs. Due to the hysteresis in a
ferroelectric matrix, the resistivity of GFEs has two
states depending on the history for any external electric field. Figure~\ref{Fig_E} shows the behavior
of the GFE conductivity on the external electric field with two distinctive features:

i) The first feature is the metal-insulator transition appearing for increasing electric field.
For weak external field the GFE is an insulator since all electronic states are localized
due to Coulomb blockade. At strong external electric field electrons can overcome the Coulomb gap moving
the GFE into a metallic state. The transition between these two states occurs for the electric field $E = E^D$.
Figure~\ref{Fig_E} shows the transition between activation and metallic regimes for temperature $T = 350K$.

ii)The second feature is the hysteresis behavior.
The most striking manifestation of the hysteresis is the
strong dependence of the transition field $E^D$ on the state of a ferroelectric matrix.

We introduce the fields corresponding to different hysteresis branches as $E^D_{>,<}$.
The difference between these fields is controlled by the internal parameters of the system.
One can distinguish two different situations.
The curves shown in Fig.~\ref{Fig_E}(a) correspond to the situation when the
average internal field $E_i=5.7\cdot10^7$ V/m is smaller than the FE switching field $E_s=6.3\cdot10^7$ V/m.
In this limit the effect is pronounced. In the opposite limit, $E_i > E_s$,
there is no difference between $E^D_<$ and $E^D_>$, see Fig.~\ref{Fig_E}(b)).
Figures~\ref{Fig_E}(a) and~\ref{Fig_E}(b)) are plotted for the following set of parameters:
the tunneling conductance $g^t_0=0.2$, the grain size $a=5$ nm,
$\Omega/\Omega_{fe}=1.5$, parameters $\alpha$ and $\beta$ are chosen to get
the above mentioned switching fields, and $\zeta=10^{-7}$.

\begin{figure}[t]
\includegraphics[width=3.25in, keepaspectratio=true]{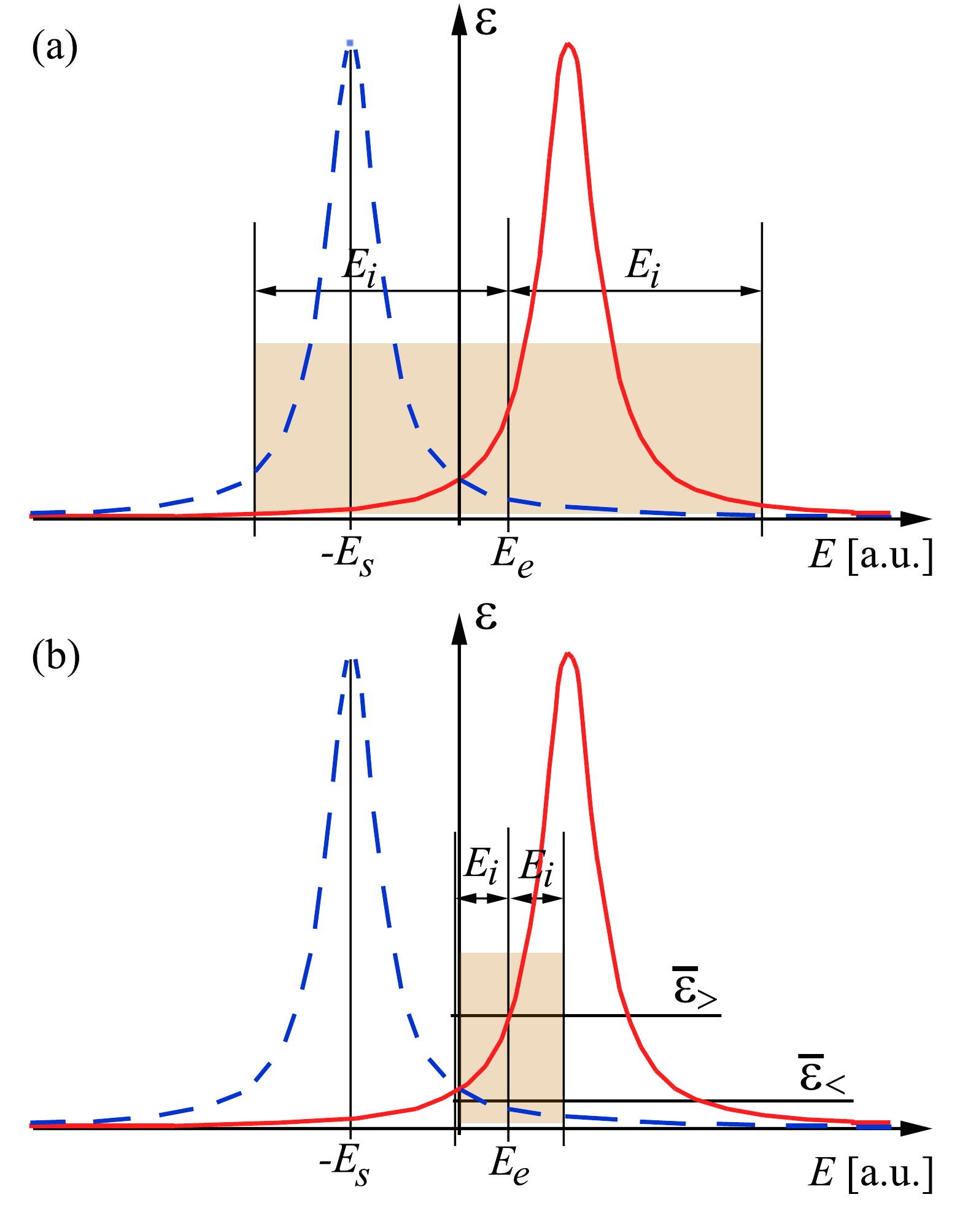}%
\caption{\label{Fig_LocPer} (color online) The local permittivity $\epsilon$ of the FE matrix vs. local electric field, $E=E_e + E_i$, with
$E_e$ and $E_i$ being the external and internal electric fields, respectively.
Solid and dash lines correspond to two hysteresis branches. According to the procedure described in the Sec.~\ref{sec:model} the dielectric permittivity of the whole GFE is the average of the local permittivity over different internal field orientations. The filled area shows the region of averaging.
There are two different situations: (a) The internal field $E_i$ is larger than the switching field $E_s$ and
(b) The internal field $E_i$ is less than the switching field $E_s$.}
\end{figure}

The transition field $E^D=e^2/(a^2\epsilon)$ is determined by the average dielectric permittivity, $\epsilon$.
Thus to understand the two limits mentioned above
one has to study the dependence of the GFE dielectric permittivity $\epsilon$ on the external electric field, $E_e$.
Figure~\ref{Fig_LocPer} shows the dependence of the local FE permittivity $\epsilon$ on the
local electric field consisting of internal field $\vec{E}_i$ and external field $\vec{E}_e$.
The two curves correspond to the two hysteresis branches.
The permittivity of the whole GFE can be found by averaging over all orientations of the internal field, see Sec.~\ref{sec:model}.
The local susceptibility should be averaged over the
field interval [$E_e-E_i,E_e+E_i$]. If the internal field $E_i\gg E_s$, see Fig.~\ref{Fig_LocPer}(a),
then the averaging produces the same result for both branches and
one gets the same dielectric permittivity unless the external
electric field is less than $E_i-E_s$. Therefore if the transition field
$E^D < E_i-E_s$ there is no difference between $E^D_<$ and $E^D_>$, see Fig.~\ref{Fig_E}(b).
If $E_i < E_s$, see Fig.~\ref{Fig_LocPer}(b), then the averaged dielectric permittivity is
different for the two branches at any finite external field. Therefore in this case $E^D_<\ne E^D_>$.

\subsection{Comparison with experiment}

Here we compare our results with available experimental data on electron transport in composite ferroelectrics~\cite{Dai2013}.
Experimentally, the current voltage characteristics has two peculiarities [see Fig. 3(a) of Ref.~\cite{Dai2013}]: i) switching of the
resistivity at a certain voltage and ii) a current voltage hysteresis effect.
As can be seen in Fig.~\ref{Fig_E}, the same features are present in our model:
i) The current jump appears due to a transition from the insulating phase with cotunneling transport mechanism to a metallic phase
with suppressed Coulomb blockade. ii) The memory effect appears due to a ferroelectric matrix hysteresis.
Thus, the data of Ref.~\cite{Dai2013} can be qualitatively described by our theory.

We note, that our Fig.~\ref{Fig_E} shows bipolar switching behavior
in contrast to the unipolar switching mechanism reported in Ref.~\cite{Dai2013}. This difference is related to the fact
that we assume an infinite relaxation time of the polarization of the FE matrix here.
However, if the relaxation time is comparable to the time the loop is traversed,
unipolar switching is possible as well.

We note, that the variation of the switching voltage for different hysteresis loops observed in Ref.~\cite{Dai2013} is an effect, which cannot be described by the framework presented here.
The switching of the resistance appears when the Coulomb blockade
is suppressed by an external field along a single conductive chain.
The first conductive chain is determined by the current distribution
of the electrons in the metal particles and impurities.
Therefore it can be different for different
sweeping loops and so does the switching voltage. In our consideration we average the
current over a large system size smearing out the charge distribution fluctuations.
Therefore the switching voltage is time independent.
This is not the case in Ref.~\cite{Dai2013}, since the thickness of the GFE in their experiment is rather small.

We also mention that current-voltage hysteresis loops were observed in granular metals~\cite{Cho2010}, i.e.,
in systems consisting of metallic grains embedded into a simple insulator.
In this case memory effects can be understood using the Simmons-Verderber model~\cite{Verder1967},
where electrons are trapped by defects inside the insulator (the metal particles in the case of granular metals) and stored in these defects for long times. This modifies the potential profile for electrons moving
through the system and changes the resistance. This model can be also used for a
description of current voltage hysteresis in GFEs.
In order to discriminate between these two effect on can heat the system above the ferroelectic Curie temperature,
such that the contribution of the hysteresis due to the FE matrix can be neglected.

\subsection{Influence of the FE matrix on the electron transport in the metallic regime}

\begin{figure}[t]
\includegraphics[width=3.25in, keepaspectratio=true]{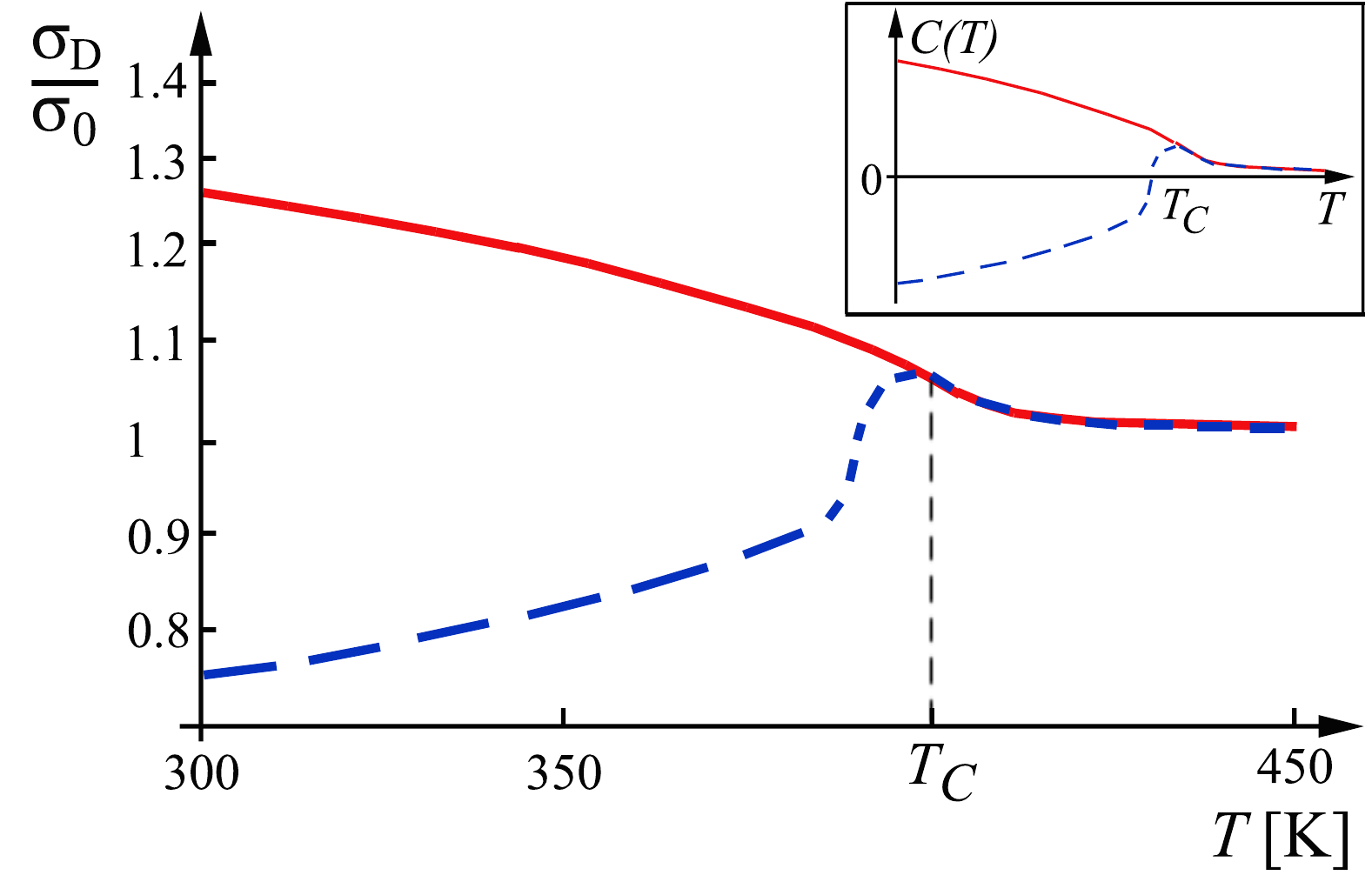}%
\caption{\label{Fig_Met} (color online) Conductivity of a GFE vs. temperature in the metallic regime, Eq.~(\ref{24}), at fixed external
electric field $E_e=3\cdot10^7$ V/m. Solid and dash lines correspond to the two hysteresis branches.
The inset shows the behavior of the correlation function $C(T)$ on temperature, Eq.~(\ref{Eq_14}).
The conductivity in the metallic regime is controlled by the correlation function.}
\end{figure}

In the metallic regime the dielectric permittivity of the FE matrix does not
play an important role on the electron transport of GFEs. However, it influences
the tunneling conductance between grains. In this region the correlation function $C$
of local electric polarization and the local electric field becomes important.
Figure~\ref{Fig_Met} shows the behavior of
the conductivity in the metallic region (external field $E_e=3\cdot10^7$ V/m) vs. temperature.
The parameters of the GFE are chosen to be the same as for the transport phase diagram
except for parameter $\zeta$, which is now larger $10^{-7}$.
The presence of the FE matrix leads to the occurrence of two resistive states. The
upper branch corresponds to the case of local polarization $P$ of FE matrix co-directed
with the external electric field $E_e$. In this case the correlation function is positive, $C>0$ and
the intergrain tunneling conductance and thus the conductivity increase.
The lower branch corresponds to the case of local polarization $P$ counter-directed to the external field
(due to the hysteresis phenomenon) leading to the negative correlation function $C<0$.
In this case the intergrain tunneling conductance decreases resulting in the decrease
of conductivity $\sigma_D$.
For temperatures $T > T_{\Cc}=400$ K  the memory effects are absent leading to a single resistive state.

\section{Conclusion}\label{sec:Conclusion}

We investigated the electron transport in composite ferroelectrics consisting of metallic grains
embedded in a ferroelectric matrix and show that depending on the external electric field and temperature
three transport regimes are possible: 1) multiple electron cotunneling, 2) sequential tunneling,
and 3) metallic transport. We showed that the crossover between different regimes can be studied by changing the
temperature or the external electric field leading to a strongly non-linear conductivity behavior
and large conductivity jumps. The microscopic reason for the crossover between different regimes
is the changing of the Coulomb gap due to the variation of dielectric permittivity of the
ferroelectric matrix under the influence of temperature or electric field.
This interesting effect arises due to the interference of granular morphology and ferroelectric matrix.

Another peculiarity of electron transport in composite ferroelectrics occurs due to the hysteretic behavior of the ferroelectric matrix. It leads to the existence of two different intermediate states with different average electrical polarization and correlation function of microscopic electric field and microscopic polarization. These two states have different conductivity.

We showed that our theory is in qualitative agreement with recent experiments on transport properties of granular ferroelectrics.

In addition, we show that the main parameters determining the transport in composite ferroelectrics
are: 1) the correlation function of intrinsic microscopic field and the local electric
polarization and 2) the dielectric permittivity of the ferroelectric matrix.

\section{Acknowledgments}

A.~G. was supported by the U.S. Department of Energy Office of Science under the Contract No. DE-AC02-06CH11357.
I.~B. was supported by NSF under Cooperative Agreement Award EEC-1160504.

\appendix

\section{Applicability of the model\label{ApA}}

In this appendix we consider the applicability of our method. It is based on a mean field approach, meaning that
fluctuations around the order parameter are small and therefore cannot suppress the order parameter.

First, we estimate the correlation length of electrical order parameter, $r_c$, which is given
by the expression,  $r_c=\sqrt{g/(\alpha(T-T_{\Cc}))}$, with $\alpha$ and $g$ being the constants in the expression
of the free energy density of the ferroelectric, $F = F_0+\alpha (T-T_{\Cc}) P^2+\beta P^4+g(\delta_{\vec{r}}P)^2$.
The parameters $g$ and $\alpha$ can be estimated as $\alpha=0.01$ K$^{-1}$, $g\approx 3\cdot 10^{-22}$ cm$^2$,~\cite{Frid2010rev}. In our consideration
the direction of the polarization $P$ is determined by the local anisotropy field appearing due to grain
boundaries. This assumption is valid if the ferroelectric domain wall thickness (or correlation length) is
less than the characteristic length scale of the spatial variations of the anisotropy field. The latter is of the order of the grain size ($\sim 5$ nm).
Therefore for correlation length $r_c < 5$ nm our consideration is justified. With the parameters provided above
this inequality holds for temperatures $|T-T_{\Cc}|>12 K$.

Second, the mean field theory for $3d$ samples is valid for temperatures~\cite{Landau5}
\begin{equation}\label{Ap_1}
\frac{k_B T_{\Cc} \, \overline{\chi}}{r_c^3}\ll\frac{\alpha(T-T_{\Cc})}{\beta}=2P_0^2,
\end{equation}
where $k_B$ is the Boltzman constant, $\overline{\chi}$ is the macroscopic susceptibility,
$\beta$ is the constant in the expression for the free energy density of the ferroelectric material.
To estimate the l.~h.~s. of Eq.~(\ref{Ap_1}) we use the following set of parameters,~\cite{Frid2010rev}:
$\alpha(T-T_{\Cc})\approx 1$ ($T_{\Cc}=400$, $T=300$, $\alpha=0.01$ K$^-1$), $g\approx 3\cdot 10^{-22}$ cm$^2$.
For the correlation length we find $r_c\approx 1.5$ nm. For these parameters  Eq.~(\ref{Ap_1}) is satisfied.

The electrical polarization for thin ($\sim 5$nm) films of BaTiO$_3$
is about 0.4 C/m$^2$ (1.2$\cdot 10^5$ statC/cm$^2$ in cgs),~\cite{Frid2006rev}. The critical thickness for
this material is of order of $1$nm. For the polarization $P_0\approx 3\cdot 10^5$ statC/cm$^2$ and
susceptibility $\chi\sim 1/\alpha(T-T_{\Cc})\approx 1$ we find
\begin{equation}\label{Ap_2}
\frac{k_B T_{\Cc} \overline{\chi}}{r_c^3}\approx 10^{7} \textrm{statC}^2/\textrm{cm}^4,
\end{equation}
and
\begin{equation}\label{Ap_3}
P_0^2\approx 10^{11} \textrm{statC}^2/\textrm{cm}^4.
\end{equation}
Therefore inequality (\ref{Ap_1}) is well satisfied at room temperature.

Decreasing the temperature one can effectively reduce the dimensionality of the sample.
For the correlation length larger than the ferroelectric thickness, the three dimensional condition,
Eq.~(\ref{Ap_1}), needs to be replaced by the following condition for the applicability of the mean field theory
\begin{equation}\label{Ap_4}
\frac{k_B T_{\Cc} \overline{\chi}}{r_c^2L}\ll\frac{\alpha(T-T_{\Cc})}{\beta}=2P_0^2,
\end{equation}
where $L$ is the ferroelectric thickness.
As one can see Eq.~(\ref{Ap_4}) is also satisfied for our set of parameters.
Thus the requirement of a small correlation length in comparison with the grain sizes
is the strongest restriction determining the validity of our considerations.

\section{Average characteristics of GFE \label{ApB}}

Here we discuss the average thermodynamic characteristics of GFE.
The mutual orientation of local normal $\vec{n}$, internal $\vec{E}_i$, and external $\vec{E}_e$ electric fields is random. We introduce angles ($\theta_e$, $\phi_e$) and ($\theta_i$, $\phi_i$) describing the orientation of fields $\vec{E}_e$ and $\vec{E}_i$ with respect to the local normal, $\vec{n}$. For uniform distribution of angles ($\theta_i$, $\phi_i$) the distribution function is $\omega_{i}(\theta_{i},\phi_{i})=1/(4\pi)$. The distribution function of the angles ($\theta_e$, $\phi_e$) is described by the following expression
\begin{equation}\label{Eq_4}
\omega_{e}(\theta_{e},\phi_{e})=\frac{1}{4\pi}\left\{\begin{array}{l} 1+\sgn(E_e),~ 0<\theta_e\le\pi/2, \\ 1-\sgn(E_e),~\pi/2<\theta_e\le\pi. \end{array}\right.
\end{equation}
In general, the distributions can be anisotropic for grains forming a regular array. However, here we concentrate on the isotropic case.

We now calculate the average polarization $\overline{\vec{P}}$ for finite external electric field, $\vec{E}_e$. For the isotropic model the average polarization is parallel to the external field $\overline{\vec{P}} = \overline{P} \, \vec{x}_0$. The local polarization is directed along the local normal and its projection on the positive direction is $(\vec{P}\cdot\vec{x}_0)=P(E_{\vec{n}})|\cos(\theta_e)|$. Also we assume that the magnitude of the internal field $|\vec{E}_i|$ is spatially homogeneous. The generalization for an inhomogeneous distribution of internal fields is straightforward. For the average polarization we obtain
\begin{equation}\label{Eq_6}
\begin{array}{l}\overline{P}=\iint\limits^{~~2\pi}_{0}d\phi_i d\phi_e\iint\limits^{~~\pi}_{0}\sin(\theta_i)d\theta_i\sin(\theta_e)d\theta_e \times \\ ~~~~~\times P(E_{\vec{n}})|\cos(\theta_e)| \omega_e(\theta_{e},\phi_{e}) \omega_i\,.\end{array},
\end{equation}
where $\omega_e(\theta_{e},\phi_{e})$ is the distribution function defined by Eq.~\ref{Eq_4}.
Note, that the average polarization does not enter directly into the expressions for the electron transport.

Besides the average polarization, an important characteristic is the average dielectric susceptibility $\overline{\chi}$. The composite material is isotropic in the paraelectric phase for zero external field an hence, the dielectric susceptibility $\overline{\chi}$ is also isotropic. However, for finite electric field it is necessary to distinguish the longitudinal and transverse dielectric permittivity. The anisotropy of $\overline{\chi}$ becomes important for an external field of the order of the internal field. For these fields the transport is metallic meaning that the Mott gap is vanishingly small. Thus the susceptibility is not important for strong fields. Below we consider the limit of strong internal fields, $E_e \ll E_i$, and introduce the coordinate system related to the field $\vec{x}_0$. The direction along vector $\vec{x}_0$ is denoted by the subscript $\parallel$ and the direction perpendicular to $\vec{x}_0$ is denoted by subscript $\perp$. The average longitudinal susceptibility $\overline{\chi}_{\parallel}=\partial{\overline{P}_{\parallel}}/\partial{E}_{e\parallel}$ can be calculated as follows
\begin{equation}\label{Eq_11}
\begin{array}{l}\overline{\chi}_{\parallel}=\iint\limits^{~~2\pi}_{0}d\phi_i d\phi_e\iint\limits^{~~\pi}_{0}\sin(\theta_i)d\theta_i\sin(\theta_e)d\theta_e
\times \\ ~~~~~\times \chi_{\vec{n}}(E_{\parallel})|\cos^3(\theta_e)| \omega_e(\theta_{e},\phi_{e}) \omega_i.\end{array}
\end{equation}
The average transverse susceptibility $\overline{\chi}_{\perp}=\partial{\overline{P}_{\perp}}/\partial{E}_{e\perp}$ is determined by the expression
\begin{equation}\label{Eq_12}
\begin{array}{l}\overline{\chi}_{\perp}=\iint\limits^{~~2\pi}_{0}d\phi_i d\phi_e\iint\limits^{~~\pi}_{0}\sin(\theta_i)d\theta_i\sin(\theta_e)d\theta_e \times \\ ~~~~~\times \chi_{\vec{n}}(E_{\parallel})|\cos(\theta_e)|\sin^2(\theta_e)\sin^2(\phi_e) \omega_e \omega_i.\end{array}
\end{equation}
For small external fields Eqs.~(\ref{Eq_11}) and (\ref{Eq_12}) for the susceptibility can be simplified to
\begin{equation}\label{Eq_13}
\begin{array}{l}\overline{\chi}_{\parallel}=\frac{1}{12E_i}\int\limits^{E_i}_{-E_i}\chi_{\vec{n}}(\varepsilon)d\varepsilon+\frac{E_e}{16E_i}(\chi_{\vec{n}}(E_i)-\chi_{\vec{n}}(-E_i)), \\ \overline{\chi}_{\perp}=\frac{1}{12E_i}\int\limits^{E_i}_{-E_i}\chi_{\vec{n}}(\varepsilon)d\varepsilon+\frac{E_e}{32E_i}(\chi_{\vec{n}}(E_i)-\chi_{\vec{n}}(-E_i)). \end{array}
\end{equation}
Above the Curie temperature, $T > T_{\Cc}$ the second terms of both equations are zero and therefore the
susceptibility is isotropic. It is isotropic for zero external field $E_e$.
The lowest order expansion of $\overline{\chi}$  in external field, $E_e$, given in Eqs.~(\ref{Eq_13}), has a finite linear contribution (2nd terms).
This indicates a finite remanent electric polarization at zero external field and therefore is a signature of the hysteretic behavior.

One more important characteristic quantity of composite ferroelectrics is the
correlation function of electric fields and polarization $C=\av{(\vec{E}_i+\vec{E}_e)\cdot\vec{P}}$.
It describes corrections to the tunneling conductance in polarization $\vec{P}$ and
determines the transport properties of a sample.
In contrast to the dielectric susceptibility this correlation function is important in the whole
range of external electric fields. The correlation function $C$ is given by the following expression
\begin{equation}\label{Eq_14}
\begin{array}{l}C =
\iint\limits^{~~2\pi}_{0}d\phi_i d\phi_e\iint\limits^{~~\pi}_{0}\sin(\theta_i)d\theta_i\sin(\theta_e)d\theta_e \times \\ ~~~~~\times P(E_{\vec{n}})E_{\vec{n}}|\cos(\theta_e)| \omega_e(\theta_{e},\phi_{e}) \omega_i.\end{array}
\end{equation}
Simplifying Eq.~(\ref{Eq_14}) we obtain
\begin{equation}\label{Eq_15}
\begin{array}{l} C=\frac{1}{4E_i E_e}\left\{\int\limits^{E_i}_{-E_i}P(\varepsilon)(\varepsilon+E_i)\varepsilon d\varepsilon + 2E_i\int\limits^{E_e-E_i}_{E_i}P(\varepsilon)\varepsilon d\varepsilon+ \right.\\ \left.
+\int\limits^{E_e-E_i}_{E_e+E_i}\!P(\varepsilon)(E_i-\varepsilon+E_e)\varepsilon d\varepsilon\right\},\,\hbox{if } E_e\!>2E\!_i~ \&~ E_e>0 \end{array}
\end{equation}
and
\begin{equation}\label{Eq_16}
\begin{array}{l} C=\frac{1}{4E_i E_e}\left\{\int\limits^{E_e-E_i}_{-E_i}P(\varepsilon)(\varepsilon+E_i)\varepsilon d\varepsilon+ E_e\int\limits^{E_i}_{E_e-E_i}P(\varepsilon)\varepsilon d\varepsilon+ \right.\\ \left.
+\int\limits^{E_e+E_i}_{E_i}\!P(\varepsilon)(E_i-\varepsilon+E_e)\varepsilon d\varepsilon\right\},\,\hbox{if } E_e\!\le 2E\!_i~ \&~ E_e>0, \end{array}
\end{equation}
and
\begin{equation}\label{Eq_17}
\begin{array}{l} C=\frac{-1}{4E_i E_e}\left\{\int\limits^{E_i}_{-E_i}\!\!P(\varepsilon )(\varepsilon+E_i-E_e)\varepsilon d\varepsilon +\!\!\!\!\int\limits^{-E_i}_{E_i+E_e}2P(\varepsilon )\varepsilon E_id\varepsilon+ \right.\\ \left.
+\int\limits^{E_i+E_e}_{E_e-E_i}\!P(\varepsilon)(E_i-\varepsilon)\varepsilon d\varepsilon\right\},\,\hbox{if } |E_e|\!>2E\!_i~ \&~ E_e\le 0, \end{array}
\end{equation}
and
\begin{equation}\label{Eq_18}
\begin{array}{l} C=\frac{-1}{4E_i E_e}\left\{\int\limits^{E_i}_{E_i+E_e}\!\!\!P(\varepsilon)(\varepsilon+E_i-E_e)e\varepsilon d\varepsilon- \!\!\!\!\int\limits^{E_i+E_e}_{-E_i}\!\!P(\varepsilon)\varepsilon E_ed\varepsilon+ \right.\\ \left.
+\int\limits^{-E_i}_{E_e-E_i}\!P(\varepsilon)(E_i-\varepsilon)\varepsilon d\varepsilon\right\},\,\hbox{if } |E_e|\!\le 2E\!_i~ \&~ E_e\le 0. \end{array}
\end{equation}

\bibliography{granule}

\end{document}